\documentclass[aip,graphicx,longbibliography,reprint]{revtex4-1}

\usepackage{amssymb}
\usepackage{amsmath}
\usepackage{cases}
\usepackage{graphicx,color}
\usepackage{dcolumn}
\usepackage{bm}
\usepackage{hyperref}
\usepackage{cases}

\begin{document}


\title{Spatters and Spills: Spreading Dynamics for Partially Wetting Droplets}



\author{Sylvia C. L. Durian}
\affiliation{Department of Mathematics, University of Chicago, Chicago, IL 60637}
\altaffiliation{}

\author{Sam Dillavou}
\affiliation{Department of Physics and Astronomy, University of Pennsylvania, Philadelphia, PA 19104}
\altaffiliation{}

\author{Kwame Markin}
\affiliation{Department of Engineering, Swarthmore College, Swarthmore, PA 19081}
\altaffiliation{}

\author{Adrian Portales}
\affiliation{Department of Mechanical Engineering, University of Texas Rio Grande Valley, Edinburg, TX 78539}
\altaffiliation{}

\author{Bryan O. Torres Maldonado}
\affiliation{Department of Mechanical Engineering and Applied Mechanics, University of Pennsylvania, Philadelphia, PA 19104}
\altaffiliation{}

\author{William T. M. Irvine}
\affiliation{Department of Physics, University of Chicago, Chicago, IL 60637}
\altaffiliation{}

\author{Paulo E. Arratia}
\affiliation{Department of Mechanical Engineering and Applied Mechanics, University of Pennsylvania, Philadelphia, PA 19104}
\altaffiliation{}

\author{Douglas J. Durian}
\email{djdurian@physics.upenn.edu}
\affiliation{Department of Physics and Astronomy, University of Pennsylvania, Philadelphia, PA 19104}
\altaffiliation{}

\date{\today}

\begin{abstract}
We present a solvable model inspired by dimensional analysis for the time-dependent spreading of droplets that partially wet a substrate, where the spreading eventually stops and the contact angle reaches a nonzero equilibrium value.  We separately consider small droplets driven by capillarity and large droplets driven by gravity. To explore both regimes, we first measure the equilibrium radius versus a comprehensive range of droplet volumes for four household fluids, and we compare the results with predictions based on minimizing the sum of gravitational and interfacial energies. The agreement is good, and gives a reliable measurement of an equilibrium contact angle that is consistent in both small and large droplet regimes. Next we use energy considerations to develop equations of motion for the time dependence of the spreading, in both regimes, where the driving forces are balanced against viscous drag in the bulk of the droplet and by friction at the moving contact line. Our approach leads to explicit prediction of the functional form of the spreading dynamics. It successfully describes prior data for a small capillary-driven droplet, and it fits well to new data we obtain for large gravity-driven droplets with a wide range of volumes.  While our prediction for the dynamics of small capillary-driven droplets assumes the case of thin nearly-wetting droplets, with a small contact angle, this restriction is not otherwise invoked.
\\ \\
\textit{For submission to Physics of Fluids, special issue on ``Kitchen Flows"}
\end{abstract}


\maketitle 


\section{Introduction}

\begin{figure}
\includegraphics[width=2.5in]{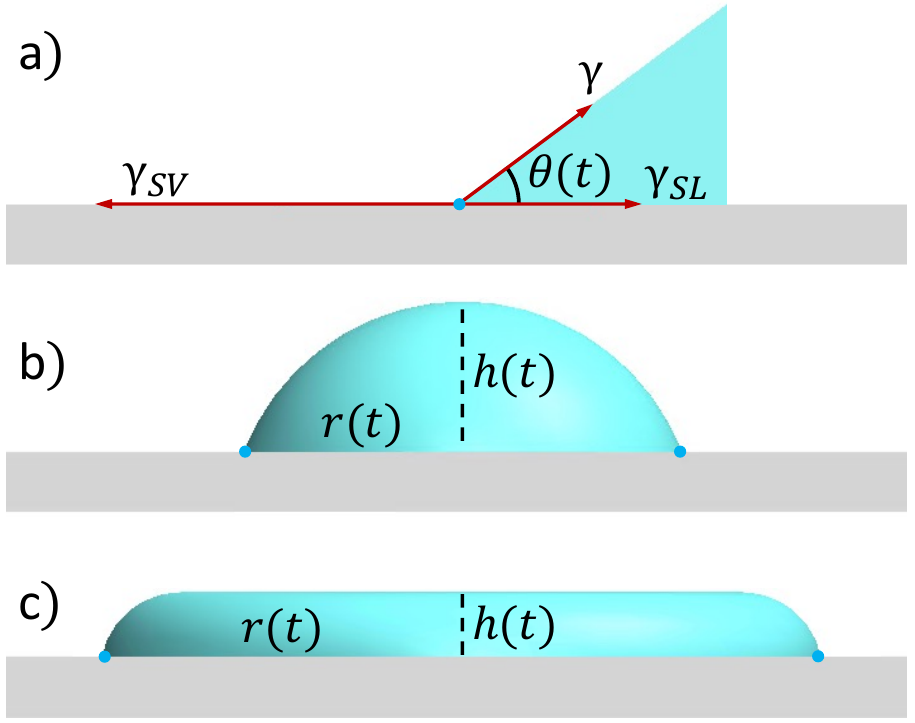}
\caption{(a) The interfaces between solid-vapor, solid-liquid, and liquid-vapor are characterized by surface tensions $\{\gamma_{SV},\gamma_{SL},\gamma\}$, respectively, specifying their energy per unit area or equivalently their tensile force per unit length.  The spreading force per length along the contact line is thus $\gamma_{SV} - \gamma_{SL} - \gamma\cos\theta = \gamma(\cos\theta_e - \cos\theta)$, where $\theta$ is the dynamic contact angle; it vanishes at the equilibrium contact angle, $\theta(t) \rightarrow \theta_e$, given by Eq.~(\ref{eq_young}). Spreading dynamics can be characterized by the time-dependent droplet radius $r(t)$ and height $h(t)$, but droplet shape depends on volume: (b) Small droplets are spherical caps, dominated by capillarity. (c) Large droplets are flat pancakes, dominated by gravity.
\label{fig_dropletsketches}}
\end{figure}

In the kitchen, liquid ingredients splash, spill, and spatter everywhere. With time, individual droplets flatten and spread but typically stop at a finite radius and a non-zero equilibrium contact angle $\theta_e$ at their outer edge (Fig.~\ref{fig_dropletsketches}). How, exactly, does the droplet height $h(t)$ decrease and the droplet radius $r(t)$ grow as a function of time toward equilibrium? This issue is widely important outside the kitchen, for the behavior of ink, adhesive, lubricant, dye, paint, etc. Spreading dynamics has been thoroughly reviewed\cite{dussan_spreading_1979, de_gennes_wetting_1985, bonn_wetting_2009} for the case of complete wetting, where $\theta_e=0$ and $r(t)\rightarrow\infty$. The growth is a power law, $r(t)\sim t^n$, but the spreading exponent $n$ can take many values depending on the dominant driving and dissipation mechanisms (see Table~II of Ref.~\onlinecite{bonn_wetting_2009}). For example, when fluid inertia and contact line friction are small, $n=1/8$ is observed for large gravity-driven droplets whereas $n=1/10$ is observed for small capillarity-driven droplets (Tanner's Law). Most of the subtlety lies in the dissipation mechanisms and behavior near the moving contact line, where special effects are required to avoid a hydrodynamics singularity\cite{huh_hydrodynamic_1971}.

For the more typical case of partial wetting, $\theta_e>0$, the dynamics of approach to equilibrium, $r(t)\rightarrow r_e$, is not discussed in the major reviews\cite{dussan_spreading_1979,  de_gennes_wetting_1985, bonn_wetting_2009}. It is, however, carefully modeled in Ref.~\onlinecite{de_ruijter_droplet_1999} for small capillary-driven droplets. The authors assume that the droplet is a spherical cap, driven by capillarity, and that dissipation is controlled by a friction coefficient at the contact line and by viscosity in the bulk. The general equations of motion must be solved numerically, but can be analyzed in different regimes. At early times, approximate power-law growth holds with a spreading exponent that is at first $n=1/7$ and then later $n=1/10$ as the dissipation changes from contact line to bulk.  At very late times, $r(t)$ is predicted to asymptote exponentially to the equilibrium value $r_e$. Other works on partial wetting dynamics considered droplet shapes, dynamic contact angles, and dissipation mechanisms\cite{degennes_dynamics_1990, brochard-wyart_dynamics_1992, eggers_contact_2005}. There has also been interest in fast dynamics at early times, when a spherical droplet initiates contact with a substrate and inertial effects are important\cite{biance_first_2004, drelich_spreading_2005, bird_short-time_2008}. However, Ref.~\onlinecite{de_ruijter_droplet_1999} remains the only work we know that models the approach of $r(t)$ toward equilibrium. It does not consider large gravity-driven droplets, and it does not report data at long enough times to observe and test their model for how $r(t)$ stops growing. We seek to fill these gaps.

Here we report experimental results obtained at home during the COVID-19 pandemic, using fluids commonly found in the kitchen. Our focus is on the growth and approach to equilibrium for the case of partially wetting Newtonian droplets of high viscosity and low inertia. Following the recipe of Lord Rayleigh\cite{rayleigh_question_1892} we cook up models to account for the observed behavior by using dimensional analyses and back-of-the-envelope calculations.

\section{Predictions}

In this section we develop approximate equations of motion that can be solved analytically for the time-dependent droplet radius, $r(t)$ versus time $t$.  To begin, we first we consider the equilibrium radius toward which $r(t)$ grows, then we make dimensional arguments for the form of the equations, and finally we use energy conservation to arrive at our main results for predicted the time-dependent spreading dynamics.

\subsection{Equilibrium Droplet Radius}

The equilibrium radius $r_e$ of a partially wetting droplet can be computed in terms of parameters that can be varied by choice of materials: the droplet volume $V$, the equilibrium contact angle $\theta_e$, and physical constants. The contact angle is set by the solid-vapor, solid-liquid, and liquid-vapor surface tensions and the Young equation
\begin{equation}
    \gamma_{SV} = \gamma_{SL} + \gamma\cos \theta_e
    \label{eq_young}
\end{equation}
representing force balance at the contact line (Fig.~\ref{fig_dropletsketches}a). If this equation is not satisfied, then the spreading force per unit length along the contact line is  $\gamma_{SV} - \gamma_{SL} - \gamma\cos\theta = \gamma(\cos\theta_e - \cos\theta)$; see Fig.~\ref{fig_dropletsketches}.

\subsubsection{Small Capillarity-Dominated Droplets}

If the droplet is small enough to neglect gravity, then it has constant internal pressure and hence forms a spherical cap (Fig.~\ref{fig_dropletsketches}b). The equilibrium droplet radius $r_e$, height $h_e$, and radius of curvature $R_e$, are determined from $V$ and $\theta_e$ by geometry alone.  The three ingredients are $V=\pi h_e(3r_e^2 + h_e^2)/6$, $\sin\theta_e=r_e/R_e$, and $R_e^2 = r_e^2 + (R_e-h_e)^2$.  These can be solved for droplet radius, height, and curvature as a function of volume and contact angle:
\begin{eqnarray}
    r_e &=& \left[  \frac{6V\cos^3\frac{\theta_e}{2} }{\pi (2+\cos\theta_e) \sin\frac{\theta_e}{2} } \right]^{1/3} \approx
    \left( \frac{4V}{\pi \theta_e} \right)^{1/3} \label{eq_resmall} \\
    h_e &=& \frac{r_e(1-\cos\theta_e)}{\sin\theta_e} \approx 
    \frac{r_e \theta_e}{2} = \left( \frac{V\theta_e^2}{2\pi} \right)^{1/3} \\
    R_e &=& \frac{r_e}{\sin\theta_e} \approx \frac{r_e}{\theta_e} = \left( \frac{4V}{\pi \theta_e^4} \right)^{1/3} \label{eq_curvaturesmall}
\end{eqnarray}
where the approximations are from an expansion for small contact angles ($\theta_e \ll 1$ radian), near complete wetting. In this limit, the droplet is a thin spherical cap and its volume is $V\approx \pi r_e^2 h_e/2$. While the full expressions in Eqs.~(\ref{eq_resmall}-\ref{eq_curvaturesmall}) are exact for all $\theta_e$, the small $\theta_e$ limit will be invoked later for the dynamics of spreading.  We note that Eq.~(\ref{eq_resmall}) gives clearly correct results for $90^\circ$ and $180^\circ$ contact angles, and its small $\theta_e$ expansion matches the expression given in the Bonn et al.\ review\cite{bonn_wetting_2009}.

\subsubsection{Large Gravity-Dominated Droplets}

If the droplet is sufficiently large, then it is shaped like a pancake (Fig.~\ref{fig_dropletsketches}c) no matter what the contact angle. Now the equilibrium height $h_e$ and spreading radius $r_e$ are determined by minimizing the total surface and gravitational potential energy
\begin{equation}
    U = (A_S - \pi r^2)\gamma_{SV} + \pi r^2 (\gamma+\gamma_{SL}) + mg(h/2)
    \label{eq_U}
\end{equation}
where $A_S$ is the area of the solid, $m=\rho V$ is the mass of the droplet, $\rho$ is its mass density, and $V=\pi r^2 h$ is its approximate volume. While $A_S - \pi r^2$ represents the area of the solid-vapor interface, the resulting $A_S\gamma_{SV}$ term is an arbitrary additive constant in the potential energy that has no influence on behavior. We neglect a small correction of order $rh\gamma_{SV}$ from the rounded edge of the droplet that would vary with time and the dynamic contact angle. With this simplification, the total potential energy versus spreading radius simplifies to
\begin{equation}
    U(r) = U_o + \pi r^2\gamma(1-\cos\theta_e) + \frac{\rho g V^2}{2\pi r^2}.
    \label{eq_UofR}
\end{equation}
Minimizing versus $r$ then gives the equilibrium radius and height as
\begin{eqnarray}
    r_e &=& \left[  \frac{(V/\lambda_c)^2}{2\pi^2(1-\cos\theta_e)}  \right]^{1/4}
            \approx \left( \frac{V/\lambda_c}{\pi \theta_e}\right)^{1/2} \label{eq_relarge} \\
    h_e &=& \frac{V}{\pi r_e^2} = \lambda_c  \sqrt{ 2(1-\cos\theta_e)} \approx \lambda_c\theta_e \label{eq_helarge}
\end{eqnarray}
where $\lambda_c = \sqrt{\gamma/(\rho g)}$ is the capillary length and the approximations are for small $\theta_e$.  Naturally enough, the droplet height is proportional to $\lambda_c$ independent of droplet volume. Note too that $r_e$ diverges and $h_e$ vanishes for complete wetting, $\theta_e=0$, for both small and large droplets. The full expressions in Eqs.~(\ref{eq_relarge}-\ref{eq_helarge}) are correct for arbitrary $\theta_e$ as long as the droplet is pancake shaped. 

\subsubsection{Crossover Volume}

The crossover volume between capillary- and gravity-dominated regimes, where Eqs.~(\ref{eq_resmall},\ref{eq_relarge}) are equal, is given by
\begin{equation}
    V_c = \frac{9\pi\lambda_c^3 \sin^6\theta_e}{2\sin^5(\theta_e/2)(2+\cos\theta_e)^2} \approx 16\pi\lambda_c^3\theta_e
    \label{eq_Vc}
\end{equation}
For nearly-wetting droplets, with $\theta_e\ll 1$~radian, note that $V_c$ is considerably smaller than the capillary volume $\lambda_c^3$, and $h_e$ is considerably less than $\lambda_c$.  See Ref.~\onlinecite{extrand2010} for an alternative approach, where the transition is identified by equating droplet height expressions from the two regimes.

Since the equilibrium radius $r_e$ scales with different powers of $V$ for large and small droplets, the values of $\{ \theta_e, \lambda_c \}$ could be reliably extracted from asymptotic analysis of $r_e$ versus $V$ data. Intermediate size droplets, and the exact droplet geometry, might be found by variational calculus, satisfying both Young's equation and pressure balance $\kappa(z)\gamma=\rho g z$ where $\kappa(z)$ is total interface curvature at distance $z$ below the top of a droplet of fixed volume.

\subsection{Dimensional Analysis}

In order for a spreading droplet to reach equilibrium, its excess gravitational and surface potential energies must be dissipated. This could be dominated by flow throughout the bulk of the droplet, as set by the viscosity $\eta$ of the fluid. Or it could be dominated by ``friction" at the moving contact line, as set by a coefficient $\zeta$ that has dimensions of viscosity and encapsulates molecular effects\cite{blake_kinetics_1969, de_ruijter_droplet_1999}.  Altogether there are nine relevant quantities $\{r, \dot r, r_e, V, \gamma, \rho, g, \eta, \zeta\}$ whose dimensions are a combination of three basis units \{g, cm, s\}.  According to the Buckingham~$\Pi$ theorem \cite{buckingham_1914}, which formalizes Rayleigh's method of dimensional analysis, one may construct $9-3=6$ independent dimensionless quantities. However, it's actually useful to consider more than just six:
\begin{eqnarray}
    \left[ \frac{\eta \dot r}{\gamma} \right] &=&
    \left[ \frac{\zeta \dot r}{\gamma} \right] =
    \left[ \frac{\eta \dot r}{\rho g r^2} \right] =
    \left[ \frac{\zeta \dot r}{\rho g r^2} \right] = 1 \label{eq_fourrdots} \\
    \left[ \frac{V}{r^3} \right ] &=&
    \left[ \frac{r}{r_e} \right ] = 1 \label{eq_twors}  \\
    \left[ \frac{\rho r \dot r}{\eta} \right] &=& \left[ \frac{\rho r \dot r}{\zeta} \right] = 1 \label{eq_twoRes}
\end{eqnarray}
In the first line, the numerators and denominators have units of force/length and represent the ratio of dissipation (in bulk or at contact line) to driving (capillarity or gravity); the first two of these are capillary numbers.  The two in the middle line represent geometrical scales. The two in the third line are Reynolds numbers, which we assume are small on approach to equilibrium. 

In practice, it often happens that only one driving mechanism and one dissipation mechanism dominate behavior. The case can be judged by the value of two other dimensionless numbers:
\begin{eqnarray}
    \textrm{Bo} &=& \frac{\rho g r^2}{\gamma} = \left(\frac{r}{\lambda_c}\right)^2 \label{eq_bond} \\
    \textrm{Co} &=& \frac{\zeta V}{\eta r^3} \label{eq_contact}
\end{eqnarray}
The first is the Bond number, which represents the ratio of gravity to capillarity.  The second is a number we have not previously encountered, which we name the ``Contact" number and which represents the ratio of dissipation at the contact line to that in the bulk.  The geometrical factor of $V/r^3$ was included in hindsight based on calculations given below; it is specific to droplets and would not apply to capillary rise, for example.

If the two Reynolds numbers are small, and if Bo and Co are both either large or small compared to 1, then three of the four numbers in Eq.~(\ref{eq_fourrdots}) are irrelevant and the one remaining number must equal $f(V/r^3)g(r/r_e)$ for some functions $f()$ and $g()$ of the numbers in Eq.~(\ref{eq_twors}).  The spirit of dimensional analysis is to guess and test reasonable forms for these function. Since $r(t)$ grows as a power law for the case of complete wetting, and since $\dot r$ must vanish at $r\rightarrow r_e$, presumably exponentially, one reasonable ansatz for the general equation of motion is
\begin{equation}
    \frac{\dot r}{v} = c\left( \frac{V}{r^3} \right)^\alpha \left[ 1 - \left( \frac{r}{r_e}\right)^\beta \right ]
    \label{eq_rdotoverv}
\end{equation}
where the characteristic speed scale $v$ is one of $\{ \gamma/\eta,\ \gamma/\zeta,\ \rho g r^2/\eta,\ \rho g r^2/\zeta \}$, and where $\{c,\ \alpha,\ \beta\}$ are numbers that may be computed or found by experiment.  At short times, when $r(t)$ is small compared to $r_e$, the equation of motion reduces to $\dot r/v=c(V/r^3)^\alpha$ and predicts power-law growth.  At long times, near equilibrium, the equation of motion reduces to $\dot r/v = c(V/r_e^3)^\alpha[\beta(r_e-r)/r_e]$ and hence predicts an exponential rise of $r(t)$ toward $r_e$. These behaviors are born out in the next sub-section, where we also compute values for the exponents.

\subsection{Equations of Motion}

Equations of motion for the growth of the droplet radius can be derived using energy conservation, by equating the rate of loss of potential energy with the power dissipated in flow.  The predictions are all of form Eq.~(\ref{eq_rdotoverv}) if only one driving mechanism and only one dissipation mechanism dominate.

For small droplets, $\mathrm{Bo}\ll 1$, the rate of surface energy loss equals the length $2\pi r$ of the contact line times the spreading force per length (see Fig.~\ref{fig_dropletsketches}) times the speed $\dot r$ of the contact line:
\begin{eqnarray}
    P_\gamma &=& 2\pi r \gamma (\cos\theta_e-\cos\theta)\dot r \\
    &\approx& \frac{16\gamma V^2}{\pi r^5}\left[1-\left(\frac{r}{r_e}\right)^6\right] \dot r
    \label{eq_Pgamma}
\end{eqnarray}
where $\theta$ is the dynamic contact angle, $\dot r=\mathrm{d}r/\mathrm{d}t$; the approximation is for small angles, and uses Eq.~(\ref{eq_resmall}).

For large droplets, $\mathrm{Bo}\gg 1$, the rate of total potential energy loss is given by differentiating Eq.~(\ref{eq_UofR}):
\begin{equation}
    P_g = -\frac{\textrm{d}U}{\textrm{d}t}
    = \frac{\rho g V^2}{\pi r^3} \left[ 1 -\left( \frac{r}{r_e} \right)^4 \right]\dot r
    \label{eq_Pg}
\end{equation}
using Eq.~(\ref{eq_relarge}) without further approximation, for any contact angle.  As for $P_\gamma$ the driving power vanishes at equilibrium. Note that the Bond number is recognized as $\mathrm{Bo} = P_g/P_\gamma$ in the limit $r\ll r_e$.

Turning now to the rate of energy dissipation, there are two known mechanisms.  The first is viscous flow throughout the bulk of the droplet with characteristic strain rate of $\dot r/h$.  For pancake-shaped droplets of arbitrary contact angle, and for spherical cap-shaped droplets with small contact angle $(\theta_e\ll 1$~radian), the height is $h \propto V/r^2$ and the viscous dissipation happens throughout the whole volume.  Therefore, the dissipated power is estimated from viscosity times characteristic strain rate squared times volume as
\begin{equation}
    P_\eta \propto \eta \left(\frac{\dot r}{h}\right)^2 V \propto \frac{\eta r^4}{V} \dot r^2
    \label{eq_Peta}
\end{equation}
The numerical proportionality factor depends on details of droplet shape and fluid flow, and has a logarithmic dependence on droplet volume compared to a length scale that arises from cutting off the flow singularity at the contact line.

The second dissipation mechanisms is friction at the contact line, which is controlled by a coefficient $\zeta$ with units of viscosity.  Per Ref.~\onlinecite{de_ruijter_droplet_1999} it is defined such that the total dissipation rate is
\begin{equation}
    P_\zeta = 2\pi r \zeta \dot r^2
    \label{eq_Pzeta}
\end{equation}
Note that Eqs.~(\ref{eq_Peta}-\ref{eq_Pzeta}) justify our definition of the Contact number as $\mathrm{Co} = P_\zeta/P_\eta = \zeta V/(\eta r^3)$. 

Equating total dissipation rate $P_\eta+P_\zeta$ to either $P_\gamma$ or $P_g$ gives two equations of motion, respectively:
\begin{eqnarray}
    \left( \frac{c_{cn}\eta r^9}{\gamma V^3} + \frac{c_{cz}\zeta r^6}{\gamma V^2} \right) \dot r
        &=& 1 - \left(\frac{r}{r_e}\right)^6 \label{eq_rdotsmall} \\
    \left( \frac{c_{gn}\eta r^7}{\rho g V^3} + \frac{c_{gz}\zeta r^4}{\rho g V^2} \right) \dot r
        &=& 1 - \left(\frac{r}{r_e}\right)^4 \label{eq_rdotbig}
\end{eqnarray}
where the numerical constants $c_{xy}$ subsume known and unknown factors in Eqs.~(\ref{eq_Pgamma}-\ref{eq_Pzeta}); these give $c_{cz}=\pi^2/8$ and $c_{gz}=2\pi^2$, but the other two are unknown at this point.  The first equation of motion is for small nearly-wetting droplets or spatters ($\textrm{Bo}\ll 1$ and small $\theta$).  The second is for large droplets or spills (any $\theta_e$).  As a first step towards solution, these can be rewritten with separated variables as
\begin{eqnarray}
{\rm d}t &=& \frac{ C_{cn}r^9 + C_{cz}r^6 }{1-(r/r_e)^6}{\rm d}r \label{eq_dtsmall} \\
{\rm d}t &=& \frac{ C_{gn}r^7 + C_{gz}r^4 }{1-(r/r_e)^4}{\rm d}r \label{eq_dtbig}
\end{eqnarray}
where the constants $C_{xy}$ all have different dimensions and depend on droplet volume. Both new expressions can be integrated for prediction of $t(r)$; however, the results (below) are a bit cumbersome and cannot be analytically inverted for $r(t)$. So we first examine early and late times, where the results are simpler:

At early times, when the denominator in Eqs.~(\ref{eq_dtsmall}-\ref{eq_dtbig}) can be ignored, the dynamics are identical to the limit $r_e\rightarrow \infty$ of complete wetting and integration gives
\begin{eqnarray}
t(r) &=& t_o + \frac{1}{10}C_{cn}(r^{10}-r_o^{10}) + \frac{1}{7}C_{cz}(r^7-r_o^7) \label{eq_sumpowsmall} \\
t(r)  &=& t_o + \frac{1}{8}C_{gn}(r^8-r_o^8) + \frac{1}{5}C_{gz}(r^5-r_o^5) \label{eq_sumpowbig}
\end{eqnarray}
where $r_o$ is the radius at time $t_o$. Note that power-law behavior holds if one dissipation mechanism dominates and the other can be ignored (but pure power-law growth hold only for $t\gg t_o$ and $r\gg r_o$; at the earliest times, the behavior is $r(t)\approx r_o+v_o(t-t_o)$ where the initial spreading speed $v_o$ depends on $r_o$ and the $C_{xy}$ values). The resulting exponents are familiar, except for the $r\sim t^{1/5}$ spreading of large gravity-driven droplets or spills with only contact line dissipation. For the other cases, the coefficients are also known:
\begin{eqnarray}
C_{cn} &=& \frac{\pi^3\ln[3V/(\pi a^3)]\eta}{12\gamma V^3} \label{eq_Ccn} \\
C_{cz} &=& \frac{\pi^2\zeta}{8\gamma V^2} \label{eq_Ccz} \\
C_{gn} &=& \frac{3^5\pi^3\eta}{2^7\rho g V^3} \label{eq_Cgn} \\
C_{gz} &=& \frac{2\pi^2\zeta}{\rho g V^2} \label{eq_Cgz}
\end{eqnarray}
The first is implied by Eq.~(33) in Ref.~\onlinecite{de_ruijter_droplet_1999}, where $a$ is a microscopic length scale introduced to cut off the hydrodynamic singularity at the moving contact line. The logarithmic factor is slightly different for the case of complete wetting, where there is a precursor film; see Eqs.~(10,69) in Ref.~\onlinecite{bonn_wetting_2009}. The second follows from $c_{cz}=\pi^2/8$, and is in accord with Eq.~(30) of Ref.~\onlinecite{de_ruijter_droplet_1999}. The third is from Eqs.~(2.29-30) in Ref.~\onlinecite{huppert_propagation_1982}. The fourth follows from $c_{gz}=2\pi^2$, but we are unaware of any precedent for this.  While three of the four individual power laws are known, it appears to be a new insight that they add per Eqs.~(\ref{eq_sumpowsmall}-\ref{eq_sumpowbig}) when both dissipation mechanisms contribute to the dynamics.

At late times, when $r_e-r$ is small, the numerator in Eqs.~(\ref{eq_dtsmall}-\ref{eq_dtbig}) is constant and the denominator expands to $1-(r/r_e)^\beta \approx \beta(r_e-r)/r_e$.  The predicted approach to equilibrium,
\begin{eqnarray}
r_e-r(t) &\propto& \exp\left[ -\frac{6t}{C_{cn}r_e^{10}+C_{cz}r_e^7}\right] \label{eq_expsmall} \\
r_e-r(t) &\propto& \exp\left[ -\frac{4t}{C_{gn}r_e^{8}+C_{gz}r_e^5}\right] \label{eq_expbig}
\end{eqnarray}
is thus exponential in time for both small and large droplets.  Note that the exponential relaxation times depend on the final spreading radius, as well as the same coefficients that appear in the early-time power laws. For small capillary-driven droplets, our result agrees with Eq.~(37) of Ref.~\onlinecite{de_ruijter_droplet_1999}. For large gravity-driven droplets, our prediction seems to be a new result.

The general solution of Eqs.~(\ref{eq_rdotsmall}-\ref{eq_rdotbig}) for small and large droplets, respectively, can be written as follows based on Eqs.~(\ref{eq_dtsmall}-\ref{eq_dtbig}):
\begin{eqnarray}
t(r) &=& t_o + C_{cn}r_e^{10}I_{96}(r) + C_{cz}r_e^7 I_{66}(r) \label{eq_gensolnsmall} \\
t(r) &=& t_o + C_{gn}r_e^{8}I_{74}(r) + C_{gz}r_e^5 I_{44}(r) \label{eq_gensolnbig} \\
I_{ij}(r) &=& \int_{r_o/r_e}^{r/r_e} \frac{x^i {\rm d}x}{1-x^j} \label{eq_Iij}
\end{eqnarray}
The $I_{ij}(r)$ are dimensionless functions of droplet radius, which we compute to be
\begin{widetext}
\begin{eqnarray}
I_{96}(r) &=& \frac{1}{12}\ln\left[  \frac{(r_e^2-r_o^2)^2(r_e^4+r_e^2r^2+r^4)}{(r_e^2-r^2)^2(r_e^4+r_e^2r_o^2+r_o^4)} \right]
+\frac{\sqrt{3}}{6}\Bigg[ \arctan\left(\frac{r_e+2r}{\sqrt{3}r_e}\right) -\arctan\left(\frac{r_e+2r_o}{\sqrt{3}r_e}\right) \nonumber \\ &~& +\arctan\left(\frac{r_e-2r}{\sqrt{3}r_e}\right) -\arctan\left(\frac{r_e-2r_o}{\sqrt{3}r_e}\right) \Bigg] - \left(\frac{r^4-r_o^4}{4r_e^4}\right)
\label{eq_I96} \\
I_{66}(r) &=& \frac{1}{12}\ln\left[ \frac{(r_e+r)^2(r_e-r_o)^2(r_e^2+r_er+r^2)(r_e^2-r_er_o+r_o^2)}{(r_e-r)^2(r_e+r_o)^2(r_e^2-r_er+r^2)(r_e^2+r_er_o+r_o^2)}\right]
+\frac{\sqrt{3}}{6}\Bigg[ \arctan\left(\frac{r_e+2r}{\sqrt{3}r_e}\right) \nonumber \\ &~& -\arctan\left(\frac{r_e+2r_o}{\sqrt{3}r_e}\right) -\arctan\left(\frac{r_e-2r}{\sqrt{3}r_e}\right) +\arctan\left(\frac{r_e-2r_o}{\sqrt{3}r_e}\right) \Bigg] - \left(\frac{r-r_o}{r_e}\right)
\label{eq_I66} \\
I_{74}(r) &=& \frac{1}{4}\ln\left( \frac{r_e^4-r_o^4}{r_e^4-r^4} \right) - \left(\frac{r^4-r_o^4}{4r_e^4}\right) \label{eq_I74} \\
I_{44}(r) &=& \frac{1}{4}\ln\left[ \frac{(r_e+r)(r_e-r_o)}{(r_e-r)(r_e+r_o)} \right] + \frac{1}{2}\left[ \arctan\left(\frac{r}{r_e}\right) - \arctan\left(\frac{r_o}{r_e}\right)\right] -\left(\frac{r-r_o}{r_e}\right) \label{eq_I44}
\end{eqnarray}
\end{widetext}
Note that these functions all vanish for $r\rightarrow r_o$ and diverge logarithmically for $r\rightarrow r_e$ as expected.


\section{Experiments}

We now test the above predictions, first using digitized data from a prior publication on the capillary-driven spreading dynamics for a small partial-wetting droplet \cite{de_ruijter_droplet_1999}, and then using data from our at-home experiments.  For the latter, we first describe the materials and methods, then we test behavior for (1) equilibrium radius versus droplet volume and for (2) radius versus time spreading dynamics of large gravity-driven droplets.

\subsection{de Ruijter et al. 1999}
\begin{figure}
\includegraphics[width=3.0in]{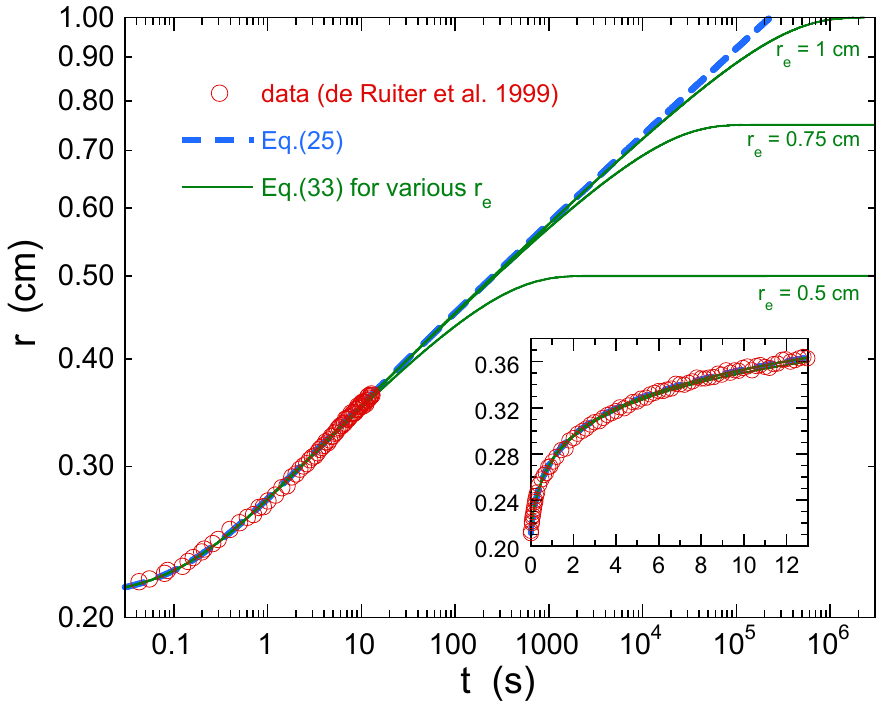}
\caption{Radius versus time for the capillary-driven spreading of a partial wetting droplet, digitized from Fig.~3 of Ref.~\onlinecite{de_ruijter_droplet_1999}.  The inset zooms into a portion of the main plot, but on linear axes.  The dashed curve is a fit to Eq.~(\ref{eq_sumpowsmall}) where droplet volume $V$ was adjusted and all other physical parameters were taken as published in Ref.~\onlinecite{de_ruijter_droplet_1999}.  The solid curves are Eq.~(\ref{eq_gensolnsmall}), using these parameter and the fitted value of $V$, for three assumed values of the final spreading radius $r_e$. }
\label{fig_deRuijterFits}
\end{figure}

The only prior radius versus time data for partial wetting droplets that we can locate is in Fig.~3 of the 1999 publication by de~Ruijter~{\it et al.}\cite{de_ruijter_droplet_1999} To compare with our predictions, we first extract their data points using WebPlotDigitizer (version 4.4, currently available at https://apps.automeris.io/wpd).  The authors provide values for $\gamma=34.3$~mN/m, $\eta=19.6$~mPa-s, $\zeta=130\eta$, and $a=1.4\mu$m, where they obtained the latter two and $\theta_e=0.10$~degrees by fit to numerical integration of their model of capillary-driven dynamics; however, they do not specify the droplet volume. So we plotted their data as $t$ versus $r$ and fit to our Eq.~(\ref{eq_sumpowsmall}) by adjusting $V$ at fixed $t_o=0$~s and $r_o=0.21$~cm.  This gives $V=5.2$~mm$^3$, in accord with the range 5.0--9.5~mm$^3$ they state in a prior publication \cite{de_ruijter_contact_1997}.  The data and our fit are in excellent agreement, as shown in Fig.~\ref{fig_deRuijterFits}. Thus our simple model works as well as the more complex one requiring numerical integration. Unfortunately, their data collection stopped well before the equilibrium spreading radius and contact angle were reached. We don't understand how they fit for $\theta_e=0.10$~degrees, since our own fits are insensitive to the contact angle over the range of their data; however, their value and $V=5.2$~mm$^3$ predict $r_e=1.5$~cm using Eq.~(\ref{eq_resmall}). Fig.~\ref{fig_deRuijterFits} also shows Eq.~(\ref{eq_gensolnsmall}) for assumed values of $\{0.5, 0.75, 1\}$~cm for $r_e$. As seen, this choice does not affect the agreement of theory and data, which is evidently in the early-time regime where Eq.~(\ref{eq_sumpowsmall}) holds independent of the equilibrium contact angle and the resulting equilibrium radius.

\subsection{Materials and Methods}

For our experiments we chose four readily available fluids that are Newtonian, safe to use at home where our experiments are performed, and have a wide range of viscosities: glycerol (Eisen-Golden Laboratories, 99.7\% anhydrous, ACS grade), blackstrap molasses (Golden Barrel, unsulfered blackstrap molasses), corn syrup (Karo, light corn syrup with real vanilla), and honey (Nature Nate's, 100\% California pure raw \& unfiltered honey).  To aid visualization, blue food coloring dye is added to all but molasses (McCormick; 5--10 drops per 50~mL of fluid). Physical properties are measured as follows and collected in Table~\ref{table_fluids}.  The mass density $\rho$ is obtained by weighing 50~mL of fluid in a graduated cylinder. The shear viscosity $\eta$ is measured using a stress-controlled rheometer (TA Instruments, DHR~30) in a 50~mm diameter cone-and-plate geometry. A custom-made solvent trap is used to avoid solvent evaporation. Newtonian behavior is found for all fluids over the full range of strain rates tested (0.1--100/s).  The liquid-vapor surface tension $\gamma$ is obtained using a 15~minute pendant drop test with a 0.9~mm needle diameter and droplet volumes between 7 and 11~$\mu$L (KSV Instruments, Attension Theta). Surface tensions are not sensitive to dye at the concentrations used, but do exhibit systematic drift during the tests; thus, we take a straight average over all times and use the full range as an estimate of uncertainty. Surface tension and density results are used to compute the capillary length for each fluid, $\lambda_c=\sqrt{\gamma/(\rho g)}$, also given in the table.

\begin{widetext}
\begin{table*}[t]
\caption{\label{table_fluids} Measured values of mass density $\rho$, dynamic viscosity $\eta$, and liquid-vapor surface tension $\gamma$; computed capillary lengths $\lambda_c=\sqrt{\gamma/(\rho g)}$; fit values for the equilibrium contact angle $\theta_e$, for four fluids.  Note that they have a wide range of viscosities, but are otherwise comparable.}
\begin{ruledtabular}
\small{\begin{tabular}{lccccc}
Fluid & $\rho$ (g/cm$^3$) & $\eta$ (g/cm$\cdot$s) & $\gamma$ (g/s$^2$) & $\lambda_c$ (cm) & $\theta_e$ (deg) \\ \hline
Glycerol & $1.250\pm0.005$ & $6.32\pm0.05$ & $62\pm5$ & $0.225\pm0.009$ & $41\pm2$ \\
Molasses & $1.404\pm0.005$ & $31.7\pm0.5$ & $63\pm2$ & $0.214\pm0.004$ & $42\pm2$ \\
Corn Syrup & $1.397\pm0.005$  & $103\pm5$ & $81\pm6$ & $0.243\pm0.009$ & $58\pm2$ \\
Honey & $1.440\pm0.005$ & $213\pm5$ & $71\pm4$ & $0.224\pm0.006$ & $48\pm 2$ \\
\end{tabular}}
\end{ruledtabular}
\end{table*}
\end{widetext}

\begin{figure}
\includegraphics[width=3.00in]{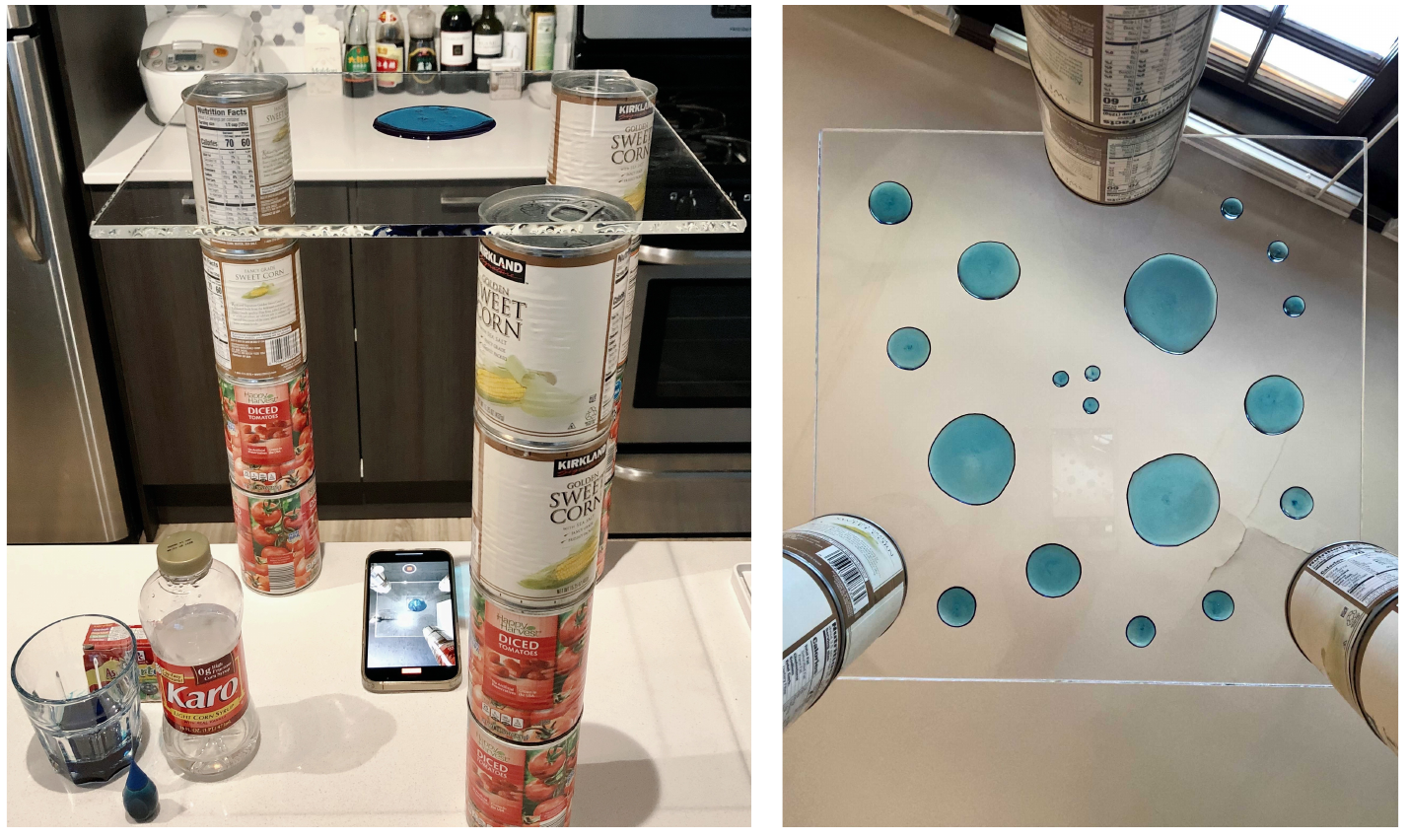}
\caption{Left: One of our setups, in the kitchen, with a large droplet of corn syrup dyed blue. Right: Example image showing such several droplets.}
\label{fig_setup}
\end{figure}

For a solid substrate, we chose a $12\times12$ square-inch borosilicate glass sheet, 1/4~inch thick (McMaster-Carr), cleaned with dishwashing detergent and dried with paper towel.  It is supported about 12~inches above a table, and made horizontal with aid of a 2~inch diameter bullseye bubble level. A photograph of one of our setups is shown in Fig.~\ref{fig_setup}.

Droplets of controlled volume between 0.02 and 20~mL are gently deposited using 1~mL or 10~mL plastic syringes.  This can take up to several seconds, during which the droplets begin spreading. Time zero is defined as the end of deposition. The droplets are illuminated from above and photographed or recorded by video from below using a smart phone in selfie mode lying on the table. Images are imported into ImageJ (NIH) for measurement of droplet area $A$, from which the effective radius is computed as $r=\sqrt{A/\pi}$.  The larger droplets are not perfectly circular, but have a smooth convex shape; the difference in major and minor axes is no more than 10\%.

\subsection{Equilibrium radius versus volume}

\begin{figure}
\includegraphics[width=3.25in]{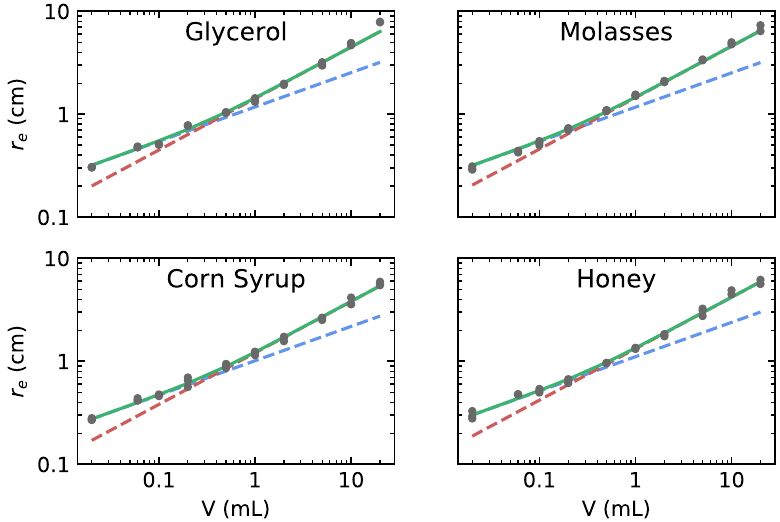}
\caption{Equilibrium radius versus droplet volume for four fluids. The solid curves represent fit to $r_e=(r_{small}^m + r_{large}^m)^{1/m}$, where small and large droplet radii expressions are given respectively by Eqs.~(\ref{eq_resmall},\ref{eq_relarge}), the crossover exponent $m$ is taken as 10, the capillary length $\lambda_c$ is taken from Table~\ref{table_fluids}, and the equilibrium contact angle $\theta_e$ is the only adjustable parameter.  Fitting results for $\theta_e$ are given in Table~\ref{table_fluids}.  The dashed lines represent the small and large radius asymptotes of the fits, {\it i.e.} Eqs.~(\ref{eq_resmall},\ref{eq_relarge}). These intersect at the crossover volume $V_c$ given by Eq.~(\ref{eq_Vc}).}
\label{fig_REvsV}
\end{figure}

To further characterize the four fluids and establish the ranges for capillary- and gravity-driven spreading, we first measure equilibrium radius $r_e$ versus a range of droplet volumes $V$ (0.02--20 mL, as wide a range as can be managed with the syringes and glass plate; three trials per volume). The data plotted in Fig.~\ref{fig_REvsV} show that the smallest droplets all increase as $r_e \propto V^{1/3}$, consistent with Eq.~(\ref{eq_resmall}) for capillary-driven spreading.  By contrast the largest droplets all increase as $r_e\propto V^{1/2}$, consistent with Eq.~\ref{eq_relarge} for gravity-driven spreading.  In the equations for both regimes, the equilibrium contact angle $\theta_e$ is the only unknown parameter. In order to interpolate between the two regimes, we fit all the data to a single empirical form, $r_e=(r_{small}^m + r_{large}^m)^{1/m}$ using Eqs.~(\ref{eq_resmall},\ref{eq_relarge}) for the small and large radii. Preliminary fits give crossover shape exponents $m$ ranging from 8 to 13, so we simply fix $m=10$ and then fit only for $\theta_e$.  The final fits, displayed in Fig.~\ref{fig_REvsV}, are excellent and hence give reliable values for the equilibrium contact angles (listed in Table~\ref{table_fluids}).  These are large enough for rough visual confirmation, similar to Figs.~\ref{fig_dropletsketches},\ref{fig_setup}.  The crossover is quite sharp and easy to identify in Fig.~\ref{fig_REvsV} as the intersection of Eqs.~(\ref{eq_resmall},\ref{eq_relarge}), given by the crossover volume of Eq.~(\ref{eq_Vc}). In particular, droplets smaller than about $V_c\approx 0.3$~mL are capillary-driven while droplets larger than $V_c$ are gravity-driven. While our procedure to measure $\lambda_c$ and fit for $\theta_e$ provides a stringent test, future researchers could use our equations to fit for both $\theta_e$ and $\lambda_c$ provided they measure a similarly wide range of droplet volumes on both sides of $V_c$.


\subsection{Radius versus time for gravity-driven droplets}

\begin{figure}
\includegraphics[width=3.0in]{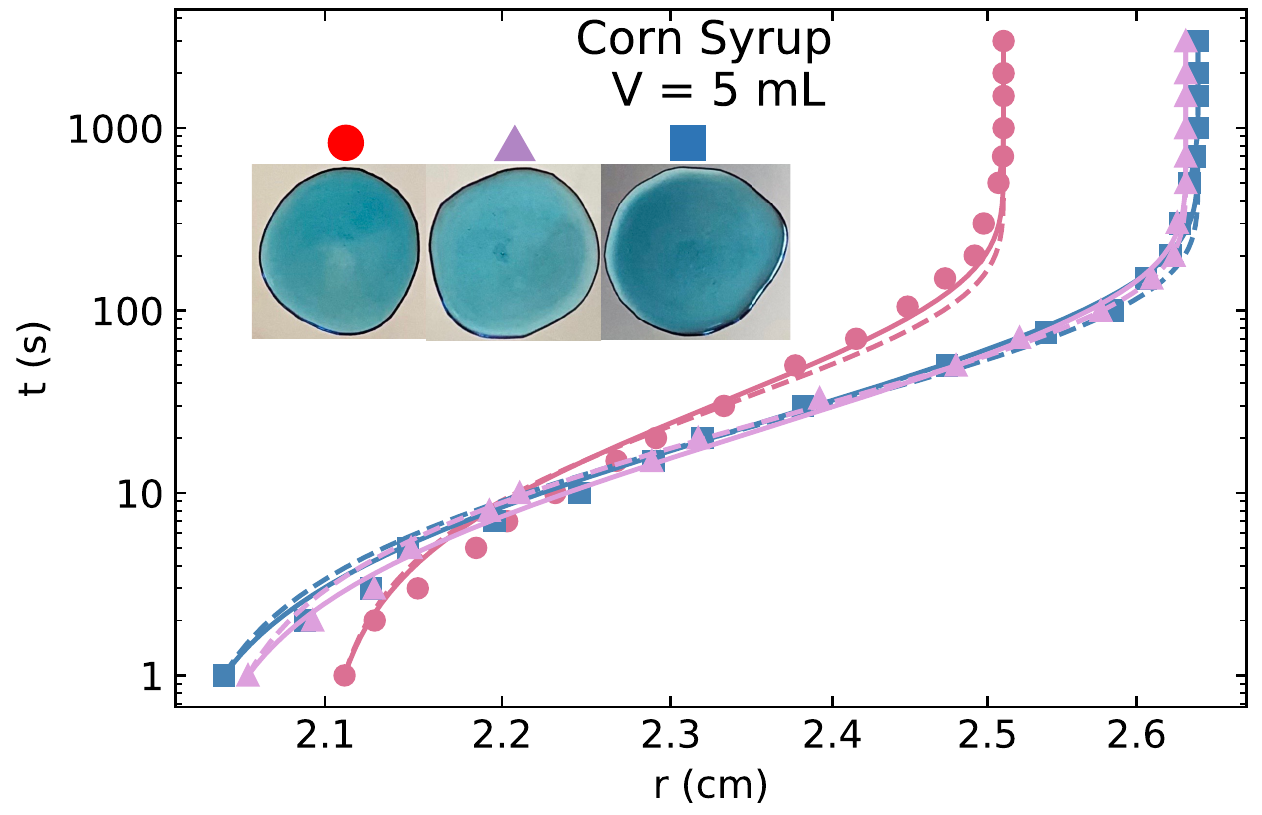}
\caption{Time versus radius for three different 5~mL droplets of corn syrup (with images taken at $t=3000$~s). Note that we plot $t$ versus $r$, rather than $r$ versus $t$, because our predictions for $t(r)$ cannnot be inverted for $r(t)$. The solid curves are fits to Eq.~(\ref{eq_gensolnbig}), where $\{t_o, r_o,r_e\}$ are taken as the measured endpoints, $C_{gz}$ is set to zero, and $C_{gn}$ is the only fitting parameter.  The dashed curves are also fits to Eq.~(\ref{eq_gensolnbig}) with $\{t_o, r_o,r_e\}$ taken as the measured endpoints, but now where $C_{gn}$ is taken by Eq.~(\ref{eq_Cgn}) and $C_{gz}$ is the only fitting parameter.}
\label{fig_dynamics}
\end{figure}

Now we are ready for our main task: Analysis of the dynamics of spreading for the partially wetting droplets.  For this, we focus on corn syrup with droplet volumes of \{0.5, 1, 2, 5, 10, 20\}~mL, all larger than $V_c\approx 0.3$~mL and hence in the gravity-driven regime. Since our four fluids are similar except for viscosities, we chose the one that is most convenient in terms of handling and time scales needed to reach equilibrium. For each volume, we conduct three trials and extract radius versus time data as described above.  Example data are shown in Fig.~\ref{fig_dynamics} for the three $V=5$~mL trials, plotted as time versus radius as per our Eq.~(\ref{eq_gensolnbig}) prediction.  The first data points are collected at 1~s after the end of droplet production, and we observe that the final equilibrium radius is reached several hundred seconds later.  Due to spreading during deposition, droplets are all within 25\% of their final value at time 1~s.  We also observe about 5\% variation in the value of the final radius, consistent with the scatter seen in Fig.~\ref{fig_REvsV}. 

The dynamics data may now be fit to our Eq.~(\ref{eq_gensolnbig}) prediction, where the $C_{gn}$ and $C_{gz}$ coefficients for bulk viscous drag and contact line friction are the only unknowns that can be used as fitting parameters. If both are adjusted, then often $C_{gz}$ is driven negative, which is unphysical; therefore, the best fits are usually obtained by adjusting only $C_{gn}$ with fixed $C_{gz}=0$. This is consistent with the fact that contact line friction would be most important at early stages, as for the case of capillary driving\cite{de_ruijter_droplet_1999}, and is also consistent with the treatment of Ref.~\onlinecite{huppert_propagation_1982} where it was not considered. The resulting fits are good, as seen by the solid curves in Fig.~\ref{fig_dynamics}, and are representative of fit obtained for the other droplet sizes.

\begin{figure}
\includegraphics[width=3.0in]{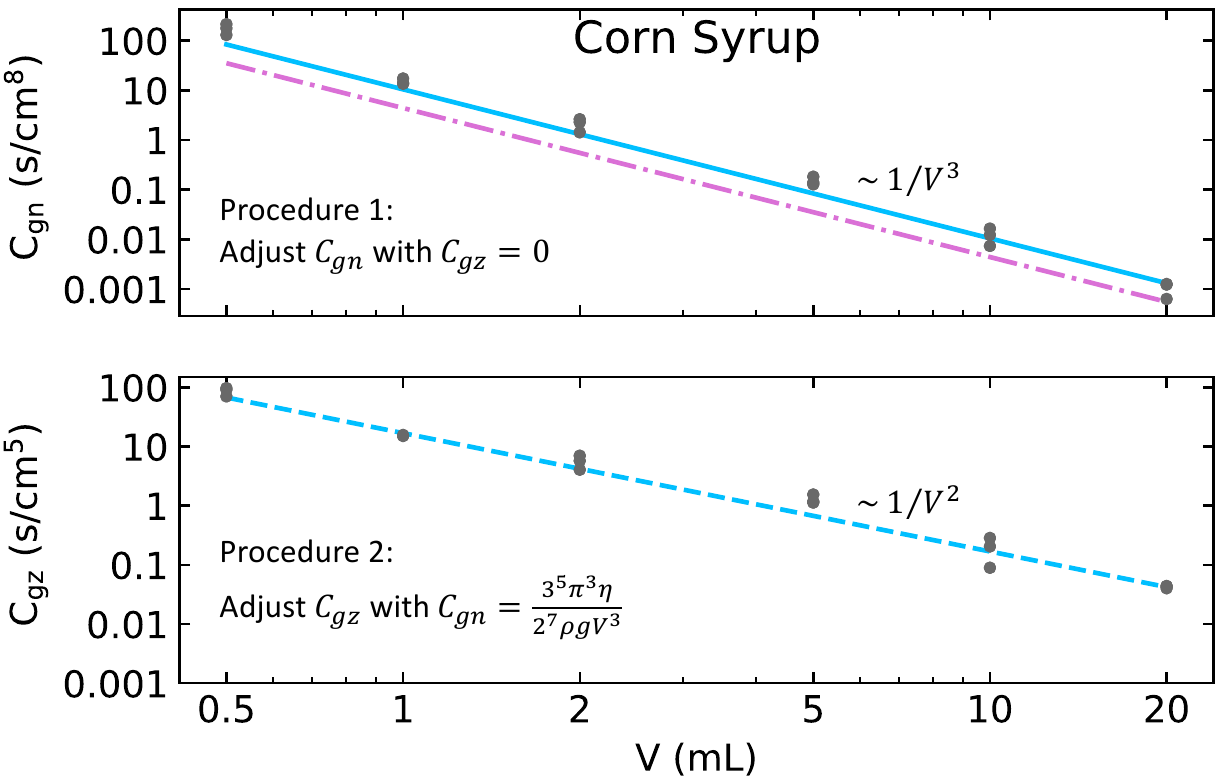}
\caption{Fitting coefficients versus droplet volume, for two different procedures of fitting Eq.~(\ref{eq_gensolnbig}) to $t(r)$ versus $r$ data on the spreading of corn syrup droplets. \emph{Top plot / first procedure}: $C_{gn}$ coefficients versus droplet volume obtained from fits such as shown in Fig.~\protect{\ref{fig_dynamics}} where $C_{gz}$ is fixed to zero. The solid line is a fit to $C_{gn} \propto \eta/(\rho g V^3)$, as expected from Eq.~(\ref{eq_Cgn}), where $\{\eta,\rho\}$ are taken from Table~\ref{table_fluids} and the numerical prefactor is found to be $140\pm40$.  For comparison, the dot-dashed line represents Eq.~(\ref{eq_Cgn}) exactly, with numerical prefactor of $3^5 \pi^3 / 2^7 \approx 58.86$ as predicted\cite{huppert_propagation_1982} for the case of complete wetting.
\emph{Bottom plot / second procedure}: $C_{gz}$ coefficients, obtained from fits the the dynamics data where $C_{gn}=3^5\pi^3\eta/(2^7\rho g V^3)$ is fixed according to Ref~\onlinecite{huppert_propagation_1982} quoted in Eq.~(\ref{eq_Cgn}). The dashed line is a fit to Eq.~(\ref{eq_Cgz}), which gives $\zeta=1100\pm300$~g/(cm$\cdot$s).}
\label{fig_cgncgz}
\end{figure}

The fitted values of the $C_{gn}$ coefficients, for fits of $t(r)$ versus $r$ to Eq.~(\ref{eq_gensolnbig}) with $C_{gz}$ fixed to zero, are plotted versus droplet volume in the top plot of Fig.~\ref{fig_cgncgz}.  There, the Ref.~\onlinecite{huppert_propagation_1982} expectation of Eq.~(\ref{eq_Cgn}) for the case of complete wetting is shown as a dot-dashed line. It lies below but parallel to the data, which are reasonably well described by $C_{gn}=(140\pm40)\eta/(\rho g V^3)$, which is like Eq.~(\ref{eq_Cgn}) but with a numerical prefactor that is a little more than twice as large. While it is encouraging that we find the expected $C_{gn}\propto 1/V^3$ scaling, we can only speculate on the discrepancy in the prefactor.  One possibility is that contact line friction is not negligible and that our fits compensate by returning an artificially large $C_{gn}$ coefficient. This is considered below, but the quality of fits and the observed $C_{gn}\propto 1/V^3$ scaling argue against it.  Another possibility is that the numerical prefactor in the $C_{gn}$ expression depends on details of the hydrodynamic flow very near the contact line and hence varies with contact angle. Indeed, for capillary-driven spreading this is the origin of the $\ln[3V/(\pi a^3)]$ term in Eq.~(\ref{eq_Ccn}) for $C_{cn}$; hence there is good reason to suspect $C_{gn}$ could vary with contact angle and that our results would be different from the complete wetting prediction. Yet another possibility is stick-slip motion from contact line pinning. This cannot be discounted because of the slightly non-circular final shapes of the droplets.

While we favor the above analysis, for completeness we attempt an alternative fitting procedure where $C_{gz}$ is adjusted to fit the $t$ versus $r$ dynamics data while $G_{gn}$ is fixed to the completely wetting value given by Eq.~(\ref{eq_Cgn}). The resulting fits are reasonable, but not as good, as represented by the dashed curves in Fig.~\ref{fig_dynamics}. Interestingly, the fitted values of $C_{gz}$ have the expected $1/V^2$ scaling, as shown in the bottom plot of Fig.~\ref{fig_cgncgz}. Fitting to Eq.~(\ref{eq_Cgz}) gives $\zeta=1100\pm300$~g/(cm$\cdot$s). This is roughly ten times the bulk viscosity, in line with the factor of 30 found in Ref.~\onlinecite{de_ruijter_droplet_1999} for a different fluid. Based on the reasonably good fits, the $1/V^2$ scaling of $C_{gz}$ with droplet volume, and the ratio of contact line friction to bulk viscosity, we cannot rule out this alternative analysis.




\section{Conclusions}

In this paper we proposed dimensional analyses and approximate but solvable models for time-dependent spreading of partially wetting droplets toward equilibrium. These predictions for spreading dynamics appear to be novel for the case of large gravity-driven droplets, and are complementary to prior work requiring numerical solution\cite{de_ruijter_droplet_1999} for the case of small capillary-driven droplets. We also collected data for both equilibrium droplet sizes and spreading dynamics, at home in the kitchen, and found good comparison with our approximate models. These advances may be helpful for characterizing the contact angle and dissipation mechanisms of fluids and for predicting their spreading behavior in a wide variety of contexts, since issues of wetting and spreading are ubiquitous not just in the kitchen but also in industry as well as in the natural world. Our work also suggests some lines for further research. For example, it would be interesting to use a variational calculus approach to capture the crossover seen in Fig.~\ref{fig_REvsV} between small and large droplets.  Further data, with closely spaced droplet volumes around $V_c$, would be helpful in this regard. It would also be worthwhile to obtain additional spreading dynamics data, and to perform hydrodynamic calculations, that could help resolve whether contact line friction need be included for the spreading of large gravity-driven droplets. Presuming not, as suggested but not proved by one of our analysis procedures, it would be useful to elucidate the value and contact angle dependence of the numerical prefactor in $C_{gn}\propto \eta/(\rho g V^3)$. And, finally, it would also be interesting to include a mechanism for contact angle hysteresis and to explore the contraction dynamics of droplets prepared with radii greater than the equilibrium value.



%
%

%


\begin{acknowledgments}
We thank Joeseph Rosenfeld and Daeyeon Lee for assistance with the pendant drop tests,  Jo\"{e}l De~Coninck for correspondence about his papers\cite{de_ruijter_droplet_1999, de_ruijter_contact_1997}, and David Miller for helpful discussions
This work was supported by NSF grants REU-Site/DMR-1659512 and MRSEC/DMR-1720530.
\end{acknowledgments}

\bibliography{DropletSpreading}

\end{document}